\documentclass[twocolumn,showpacs,pra]{revtex4}
\usepackage{amssymb}
\usepackage{amsmath}
\usepackage{graphicx}
\usepackage{bm}
\usepackage{color}

\begin{document}

\title{Crystallographic orientation and induced potential effects in
photoelectron emission from metal surfaces by ultrashort laser pulses}
\author{C. A. R\'{\i}os Rubiano and R. Della Picca }
\affiliation{CONICET and Centro At\'omico Bariloche (CNEA) Bariloche, Argentina.}
\author{D. M. Mitnik}
\affiliation{Instituto de Astronom\'{\i}a y F\'{\i}sica del Espacio
(IAFE, CONICET-UBA), Casilla de correo 67, sucursal 28, C1428EGA
Buenos Aires, Argentina.}
\affiliation{Fac. de Ciencias Exactas y Naturales, Universidad de Buenos Aires. }
\author{V. M. Silkin}
\affiliation{Donostia International Physics Center (DIPC), 20018 San Sebasti\'{a}n,
Spain, and Depto. de F\'{\i}sica de Materiales, Facultad de Ciencias Qu\'{\i}%
micas, Universidad del Pa\'{\i}s Vasco, Apdo. 1072, 20080 San Sebasti\'an,
Spain}
\affiliation{IKERBASQUE, Basque Foundation for Science, 48011 Bilbao, Spain}
\author{ M. S. Gravielle}
\affiliation{Instituto de Astronom\'{\i}a y F\'{\i}sica del Espacio (IAFE, CONICET-UBA),
Casilla de correo 67, sucursal 28, C1428EGA Buenos Aires, Argentina.}

\begin{abstract}
The influence of the crystallographic orientation of a typical metal
surface, like aluminum, on electron emission spectra produced by grazing
incidence of ultrashort laser pulses is investigated by using the
band-structure-based-Volkov (BSB-V) approximation. The present version of
the BSB-V approach includes not only a realistic description of the surface
interaction, accounting for band structure effects, but also effects due to
the induced potential that originates from the collective response of
valence-band electrons to the external electromagnetic field. The model is
applied to evaluate differential electron emission probabilities from the
valence band of Al(100) and Al(111). For both crystallographic orientations,
the contribution of partially occupied surface electronic states and the
influence of the induced potential are separately analyzed as a function of
the laser carrier frequency. We found that the induced potential strongly
affects photoelectron emission distributions, opening a window to scrutinize
band structure effects.
\end{abstract}

\maketitle

\section{Introduction}

In the last decade photoelectron emission (PE) from metal surfaces has
received renewed attention as a result of the technological achievement of
lasers with pulse durations of the order of attoseconds, which make it
possible to study the behavior of electrons in condensed matter at their
natural temporal orders \cite%
{Cavalieri07,Miaja06,Miaja08,Saathoff08,Miaja09,Mathias09,Neppl12,Neppl15}.
Such a remarkable experimental progress needs to be accompanied by intensive
theoretical research since the underlying quantum processes involve complex
many-body mechanisms, whose complete understanding is still far from being
achieved \cite{Faraggi07, Krasovskii11, Lemell15,
Borisov13,Liao14,PazourekRew2015,Baggesen08,Jouin14}. This paper aims to contribute to
the study of PE from metal surfaces due to the interaction of ultrashort
laser pulses with valence-band electrons. In particular, the article focuses
on the influence of the crystallographic orientation of the surface on
electron emission spectra, investigating the contributions of the
surface-band structure and the induced surface potential for different
crystal faces.

To describe the PE process we make use of a time-dependent distorted-wave
method named band-structure-based-Volkov (BSB-V) approximation \cite{BSB-V}.
The BSB-V approach includes an accurate description of the electron-surface
interaction, given by the band-structure-based (BSB) model \cite{chsiss99},
while the action of the laser field on the emitted electron is represented
by means of the Volkov phase \cite{Volkov35}. The BSB model is based on the
one-dimensional pseudopotential by Chulkov \textit{et al}. \cite%
{chsiss99,Chulkov97}, \ which takes into account the electronic structure of
the surface, replicating the width and position of the projected bulk energy
gap and the surface and first image electronic states \cite%
{ecpejpc78,fibev90,justss92,neinepl92,Bran4}. In this version of the BSB-V
approximation we also incorporate the contribution of the induced surface
potential, which is generated by the dynamic response of the metal surface
to the laser field \cite{Faraggi09}. \ The induced potential is derived in a
consistent way from the unperturbed BSB electronic states by using a linear
response theory \cite{Alducin2003}.

The BSB-V approximation, including the dynamic induced contribution, is
applied to evaluate double-differential (energy- and angle- resolved) PE
distributions for two different orientations of aluminum: Al(100) and
Al(111). For both crystal faces, the influence of \ partially occupied
surface electronic states (SESs) and the induced potential are examined by
considering different parameters of the laser pulse. Our results indicate
that the induced potential plays an important role in PE spectra. It makes
visible band structures signatures in PE spectra for the resonant case
wherein the laser frequency coincides with the surface plasmon frequency, as
well as for laser pulses with high carrier frequencies.


The article is organized as follows. In Sec. II we introduce the extended
version of the BSB-V approximation, which takes into account the effect of
the induced surface potential through a Volkov-type phase. In Sec. III
results are shown and discussed, while our conclusions are summarized in
Sec. IV. Atomic units are used unless otherwise stated.

\section{Theoretical method}

Let us consider a finite laser pulse, characterized by a time-dependent
electric field $\mathbf{F}_{L}(t)$, grazingly impinging on a metal surface $S
$. As a consequence of the interaction, a valence-band electron, initially
in the state $\Phi _{i}$, is ejected above the vacuum level, reaching a
final state $\Phi _{f}$. \ Within the framework of the time-dependent
distorted wave formalism \cite{Dewanga1994}, the BSB-V transition amplitude
for the electronic transition $\Phi _{i}\rightarrow \Phi _{f}$ reads \cite%
{BSB-V}:
\begin{equation}
\mathcal{A}_{if}=-i\int_{-\infty }^{+\infty }\mathrm{d}t\,\left\langle \chi
_{f}^{(BSBV)}(\mathbf{r},t)|\mathcal{V}(\mathbf{r},t)|\Phi _{i}(\mathbf{r}%
,t)\right\rangle ,  \label{AmplitudTran}
\end{equation}%
where
\begin{equation}
\mathcal{V}(\mathbf{r},t)=\mathbf{r}\cdot \mathbf{F}_{L}(t)+V_{I}(\mathbf{r}%
,t)  \label{Nu}
\end{equation}%
is the perturbative potential at the time $t$ and $\chi _{f}^{(BSBV)}(%
\mathbf{r},t)$ is the final BSB-V distorted wave function, with $\mathbf{r}$
the position vector of the active electron. The first term of Eq. (\ref{Nu})
represents the interaction potential with the laser, expressed in the length
gauge, while the second term, $V_{I}$, denotes the induced surface potential
that is produced by electronic density fluctuations caused by the external
field. The frame of reference is placed at the position of the crystal
border, which is shifted outward with respect to the position of the topmost
atomic layer by half of the interplanar distance, with the $\mathbf{\hat{z}}$
axis being oriented normal to the surface, pointing towards the vacuum
region.

Within the BSB-V approach, the unperturbed states $\Phi _{i}$ and $\Phi _{f}$
are solutions of the Schr\"{o}dinger equation associated with the
one-dimensional electron-surface potential $V_{S}(z)$ given by Ref. \cite%
{Chulkov97}, which depends on $z$, the component of $\mathbf{r}$\
perpendicular to the surface plane.\ Hence, the states $\Phi _{i}\equiv \Phi
_{\mathbf{k}_{is},n_{i}}(\mathbf{r},t)$ and $\Phi _{f}\equiv \Phi _{\mathbf{k%
}_{fs},n_{f}}(\mathbf{r},t)$ can be expressed as%
\begin{equation}
\Phi _{\mathbf{k}_{s},n}(\mathbf{r},t)=\frac{1}{2\pi }\exp \left( i\mathbf{k}%
_{s}\cdot \mathbf{r}_{s}\right) \phi _{n}(z)e^{-iEt},  \label{FunciondeOnda}
\end{equation}%
where $\mathbf{k}_{s}$ ($\mathbf{r_{s}}$) is the component of the electron
momentum (position vector) parallel to the surface plane, $\phi _{n}(z)$ is
the one-dimensional eigenfunction \ with eigenenergy $\varepsilon _{n}$
derived from the potential $V_{S}(z)$, and $E=k_{s}^{2}$ $/2+\varepsilon
_{n} $ \ is the total electron energy.

According to the grazing incidence condition and the translational
invariance of the problem in the plane parallel to the surface, the laser
field is linearly polarized perpendicularly to the surface, that is, $%
\mathbf{F}_{L}(t)=F_{L}(t)\mathbf{\hat{z}}$, where the temporal profile of
the pulse reads:
\begin{equation}
F_{L}(t)=F_{0}\,\sin {(\omega t+\varphi )}\;\sin ^{2}{(\pi \,t/\tau ),}
\label{Ft}
\end{equation}%
for $0<t<\tau $, and vanishes at all other times. In Eq. (\ref{Ft}) $F_{0}$
represents the maximum field strength, $\omega $ is the carrier frequency, $%
\tau $ is the pulse duration, and $\varphi $ is the carrier envelope phase,
which is defined as $\varphi =(\pi -\omega \tau )/2$ for symmetric pulses.
In this work we consider laser pulses with a fixed number $N$ of full cycles
inside the envelope; then, the pulse duration is defined as $\tau =NT$, with
$T=2\pi /\omega $ the laser oscillation period.

The induced potential $V_{I}$ is evaluated from a linear response theory
based on the BSB wave functions of Eq. (\ref{FunciondeOnda}) \cite%
{Silkin2010}. \ Making use of a slab geometry to derive the one-dimensional
wave functions $\phi _{n}(z)$, the induced field $\mathbf{F}_{I}=-\nabla _{%
\mathbf{r}}V_{I}(\mathbf{r},t)$ can be nearly expressed as
\begin{equation}
\mathbf{F}_{I}(z,t)=\left\{
\begin{array}{cc}
F_{I}(t)\ \mathbf{\hat{z}\qquad } & \text{for }-d<z<0, \\
0 & \text{outside,}%
\end{array}%
\right.   \label{V_ind}
\end{equation}%
where $d$ is the width of the slab, formed by a sufficiently large number of
atomic layers of the metallic crystal. The function
\begin{equation}
F_{I}(t)=-\frac{1}{2\pi }\int_{-\infty }^{\infty }\mathrm{d}\nu \widetilde{F}%
_{L}(\nu )\,\mathrm{f}_{I}(\nu )\,e^{-i\nu t}  \label{Findu}
\end{equation}%
is the induced field inside the metal at the time $t$, with $\widetilde{F}%
_{L}(\nu )$ denoting the Fourier transform of $F_{L}(t)$ and $\mathrm{f}%
_{I}(\nu )$ being the dynamic response induced by a unitary and
monochromatic electric field of frequency $\nu $.

From Eqs. (\ref{Nu}) and (\ref{V_ind}) it is possible to build $\chi
_{f}^{(BSBV)}(\mathbf{r},t)$ by introducing the distortions of both the
external and the induced fields in the momentum distribution of the final
state $\Phi _{\mathbf{k}_{fs},n_{f}}$, by means of a Volkov-type phase \cite%
{Macri03,Faraggi09,BSB-V}. It reads:
\begin{eqnarray}
\chi _{f}^{(BSBV)}\left( \mathbf{r},t\right) &=&\Phi _{\mathbf{k}%
_{fs},n_{f}}\left( \mathbf{r}-\mathbf{\hat{z}} \ \alpha _{L}\left( t\right)
,t\right)  \notag \\
&&\times \exp \left[ iz\ A_{\text{tot}}\left( z,t\right) -i\beta _{L}(t)%
\right] ,  \label{BSB-function}
\end{eqnarray}%
where the function
\begin{equation*}
A_{\text{tot}}\left( z,t\right) =\left\{
\begin{array}{lc}
\,A_{L}\left( t\right) +\,A_{I}\left( t\right) \;\; & \text{for }-d<z<0, \\
A_{L}\left( t\right) & \text{outside,}%
\end{array}%
\right.
\end{equation*}%
represents the position-dependent total vector potential at the time $t$,
with
\begin{equation}
A_{\mu}(t)=-\int\nolimits_{+\infty }^{t}\mathrm{d}t^{\prime
}\,F_{\mu}(t^{\prime }),\quad \mu=L,I,  \label{A-j}
\end{equation}%
being the vector potentials, with incoming asymptotic conditions, associated
with the laser ($\mu=L$) and induced ($\mu=I$) fields, respectively. In
turn, the functions
\begin{equation}
\alpha _{L}(t)=\int\nolimits_{+\infty }^{t}\mathrm{d}t^{\prime
}\,A_{L}(t^{\prime }),  \label{fase-alf}
\end{equation}%
and%
\begin{equation}
\beta _{L}(t)=\frac{1}{2}\int\nolimits_{+\infty }^{t}\mathrm{d}t^{\prime
}\,[A_{L}(t^{\prime })]^{2},  \label{fase-bet}
\end{equation}%
involved in Eq. (\ref{BSB-function}), are respectively related to the quiver
amplitude and the ponderomotive energy of the laser.

Finally, by replacing Eqs. (\ref{Nu}), (\ref{FunciondeOnda}), and (\ref%
{BSB-function}) in Eq. (\ref{AmplitudTran}), the BSB-V transition amplitude,
including the induced contribution, reduces to $\mathcal{A}_{if}=\,\delta (%
\mathbf{k}_{fs}-\mathbf{k}_{is})\,a_{if}$, where the Dirac delta function
imposes the momentum conservation in the plane parallel to the surface and
\begin{equation}
a_{if}=-i\,\int_{0}^{+\infty }\mathrm{d}t\,R_{if}(t)\,e^{i\left[ \Delta
\varepsilon t+\beta _{L}(t)\right] }  \label{AmplitudeInd}
\end{equation}%
represents the one-dimensional transition amplitude, with $\Delta
\varepsilon =\varepsilon _{n_{f}}-\varepsilon _{n_{i}}$ being the energy
gained by the electron during the process. The function $R_{if}$ denotes the
form factor given by
\begin{eqnarray}
R_{if}(t) &=&\int\limits_{-\infty }^{+\infty }\mathrm{d}z\,\phi
_{n_{f}}^{\ast }\displaystyle\big(z-\alpha _{L}(t)\big)\,\phi
_{n_{i}}(z)\,g_{f}(z)\,\mathcal{V}(z,t)  \notag \\
&&\times \exp \left[ -i\,z\,A_{\text{tot}}(z,t)\right] ,
\label{RFunctionInd}
\end{eqnarray}%
where $g_{f}(z)=e^{z\Theta (-z)/\lambda _{f}}$ accounts for the stopping of
the ionized electron inside the material \cite{BSB-V}, with $\Theta $ being
the unitary Heaviside function and $\lambda _{f}=\lambda \left( E_{f}\right)
$ being the electron-mean-free path as a function of the final electron
energy $E_{f}=k_{fs}^{2}$ $/2+\varepsilon _{n_{f}}$.

Analogous to Ref. \cite{BSB-V}, the BSB-V differential probability of PE
from the surface valence band can be expressed in terms of the
one-dimensional transition amplitude of Eq. (\ref{AmplitudeInd}) as:
\begin{equation}
\frac{\mathrm{d}^{2}P}{\mathrm{d}E_{f}\mathrm{d}\Omega _{f}}=2k_{f}\ \rho
(k_{fz})\sum\limits_{n_{i}}\left\vert a_{if}\right\vert ^{2}\Theta (%
\widetilde{k}_{n_{i}}-k_{fs}),  \label{dPdEdW}
\end{equation}%
where $\Omega _{f}$ is the solid angle determined by the final electron
momentum $\mathbf{k}_{f}=\mathbf{k}_{fs}+k_{fz}\mathbf{\hat{z}}$, with $%
k_{fz}=\sqrt{2\varepsilon _{n_{f}}}$. The angle $\Omega _{f}$ is defined as $%
\ \ \Omega _{f}=(\theta _{f},\varphi _{f})$, where $\theta _{f}$ and $%
\varphi _{f}$ \ are respectively the polar and azimuthal angles, with $%
\theta _{f}$ measured with respect to the surface plane. In Eq. (\ref{dPdEdW}%
), the sum indicates the addition over all the $\phi _{n_{i}}$ states with
energies $\varepsilon _{n_{i}}$ $\leq -E_{W}$ ($E_{W}$ the function work), $%
\rho (k_{fz})$ is the density of final states $\phi _{n_{f}}$ with
perpendicular momentum $k_{fz}$, and the factor $2$ takes into account the
spin states. The Heaviside function $\Theta (\widetilde{k}_{n_{i}}-k_{fs})$
comes from the momentum conservation in the direction parallel to the
surface plane, with $\widetilde{k}_{n_{i}}=\sqrt{-2\left( \varepsilon
_{n_{i}}+E_{W}\right) }$.

\section{Results}

\label{Results}

We apply the BSB-V approximation to simulate PE distributions from the
valence band of Al(100) and Al(111). Since the ejection parallel to the
polarization vector of the laser field is expected to provide the major
contribution to the PE rate \cite{BSB-V}, in this work we only consider $\ $%
electron emission normal to the surface plane, i.e., $\theta _{f}=90\ \deg $%
. The maximum field strength was chosen as $F_{0}=10^{-3}$ a.u. (intensity $%
I_{L}=3.52\times 10^{-10}$ W/cm$^{2}$), which belongs to the perturbative
range, far from the damage threshold of the material \cite{Lemell03} .

The BSB-V differential probability was evaluated from Eq. (\ref{dPdEdW}) by
varying the carrier frequency and the duration of the laser pulse. In the
calculation, the BSB wave functions $\phi _{n}(z)$ were numerically derived
by expanding them onto a basis of plane waves, defined as
\begin{equation*}
\left\{ \exp \left[ i2\pi j(z+d/2)/D\right] ,\ j=-n_{0},..,n_{0}\right\} ,
\end{equation*}%
where $2n_{0}+1$ is the number of basis functions and $D$ \ is the unit cell
width, which acts as a normalization length. By using such an expansion in
Eq. (\ref{RFunctionInd}), the form factor $R_{if}(t)$ was reduced to a
closed form in terms of the laser and induced fields, while the numerical
integration on time involved in Eq. (\ref{AmplitudeInd}) was done with a
relative error lower than $1\%$. Moreover, taking into account that the
functions $\phi _{n_{f}}$ do not allow to distinguish electrons emitted
inside the solid from those ejected towards the vacuum region, to evaluate
the emission probability we averaged the contributions from the two
different wave functions associated with the same positive energy $%
\varepsilon _{n_{f}}$ by considering that ionized electrons emitted to the
vacuum region represent approximately a $50\%$ of the total ionized
electrons from the valence band \cite{Faraggi04, BSB-V}.

The parameters associated with the different orientations of \ aluminum are
the followings. The Al(100) surface presents a work function $E_{W}=0.161$
a.u. and an interplanar distance of $3.80$ a.u., while the corresponding BSB
wave functions $\phi _{n}(z)$ were obtained by using a basis of plane waves
with $n_{0}=220$, a unit cell width $D=342.04$ a.u., and a slab width $%
d=266.00$ a.u. (i.e., $71$ atomic layers). The Al(111) surface is
characterized by a work function $E_{W}=0.156$ a.u. and an interplanar
distance of $4.39$ a.u., and the $\phi _{n}(z)$ wave functions were
evaluated using a plane wave basis with $n_{0}=170$, $D=394.92$ a.u., and $%
d=307.16$ a.u. (i.e., $71$ atomic layers). Both faces of aluminum display
the same Fermi energy, $E_{F}=0.41$ a.u., and therefore, the same surface
plasmon frequency $\omega _{s}=0.40$ a.u., which characterizes the
collective motion of valence-band electrons. The energy-dependent electron
mean free-path $\lambda (E_{f})$ was interpolated from data corresponding to
the aluminum bulk, extracted from Ref. \cite{MFPAl}.

\begin{figure*}[tbp]
\begin{center}
\includegraphics[width=1.0\textwidth]{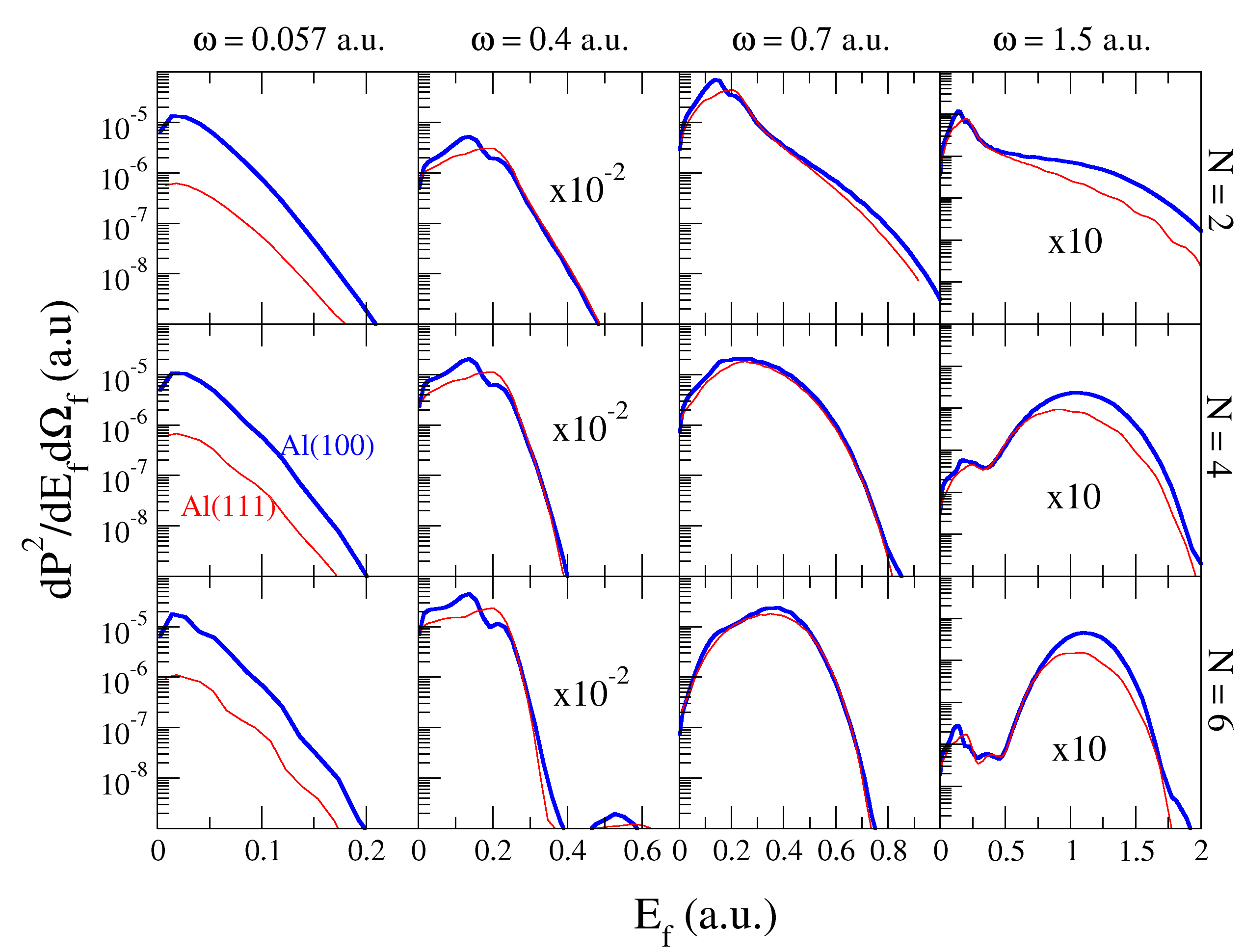}
\end{center}
\caption{(Color online) Double differential PE probabilities in the normal
direction (i.e., $\protect\theta _{f}=90\deg $), as a function of the final
electron energy $E_{f}$. Each column corresponds to a different carrier
frequency of the laser pulse: $\protect\omega =0.057$, $0.4$, $0.7$ and $1.5$
a.u. from left to right columns, respectively. Each row corresponds to a
laser pulse with a different duration, that is, with $N=2$ (upper row), $N=4$
(middle row), and $N=6$ (bottom row) cycles inside the envelope,
respectively. In all panels, BSB-V results for two aluminium faces are
displayed: Al(100), with blue thick lines, and Al(111), with red thin lines.}
\label{vsFrequency}
\end{figure*}

\bigskip

First, in order to provide an overall scenery of the influence of the
aluminum crystal face, in Fig. \ref{vsFrequency} we compare PE distributions
from Al(100) and Al(111) by considering a different number of cycles (rows)
- $N=$ $2$, $4$, and $6$ - as well as different carrier frequencies
(columns) - $\omega =0.057$, $0.4$, $0.7$, and $1.5$ a.u. For the lowest
frequency (i.e., $\omega =0.057$ a.u.), which corresponds to the
experimental value for the Ti:sapphire laser system, PE spectra from the
Al(100) surface are more than one order of magnitude higher than the ones
corresponding to the Al(111) face. But when $\omega $ increases, emission
probabilities from both aluminum faces \ become comparable in magnitude,
departing appreciably each other \ only in the high energy region for high
carrier frequencies, as observed for $\omega =1.5$ a.u. Noticeably, for the
intermediate frequency $\ \omega =0.7$ a.u. \ the electron emission
distributions from Al(100) and Al(111) are similar in the whole electron
energy range. For this frequency, the small differences between the spectra
corresponding to the different orientations for $N=2$ disappear almost completely
as $N$ increases.

Concerning the influence of the pulse duration, it is more appreciable for
high and intermediate frequencies. \ On the one hand, for short pulses, with
only two cycles inside the envelope, PE spectra (upper row of Fig. \ref%
{vsFrequency}) present a maximal emission at low electron energies, with
smoothly decreasing intensity as the velocity of ejected electrons augments.
Noteworthy, for carrier frequencies in the range $\omega $ $\gtrsim $ $%
\omega _{s}$ this low-energy maximum presents a different structure
depending on the crystal face. While for Al(111) it shows a single peak
structure, for the Al(100) face the maximum displays double-hump features,
which are particularly visible for the frequency $\omega =0.4$ a.u. resonant
with the surface plasmon frequency. In this particular case, the double
structure is clearly observed even for long pulse durations, being related
to the contribution of partially occupied SESs, as it will discussed in Sec.
III. A.

On the other hand, when the duration of the pulse increases to include
several cycles inside the envelope, the carrier frequency approximates to
the photon energy and consequently, electron distributions become governed
by the multiphoton mechanism associated with a Keldysh parameter $\gamma =$%
\ $\omega \sqrt{E_{W}}/F_{0}$ greater than the unity. For the higher
frequencies - $\omega =$ $0.4$, $0.7$ and $1.5$ a.u. - the spectra of Fig. %
\ref{vsFrequency} corresponding to $4$- and $6$-cycle laser pulses display a
broad maximum due to the absorption of one-photon of energy $\omega $, which
corresponds to the first of the above-threshold-ionization (ATI) peaks. This
ATI maximum is roughly placed at $E_{f}\simeq $ $\left\langle
E_{i}\right\rangle -U_{p}+\omega $, where $\left\langle E_{i}\right\rangle $
is the initial energy averaged over all initial states, with $\left\langle
E_{i}\right\rangle \simeq $ $-0.44$ and $-0.43$ a.u. for Al(100) and Al(111)
respectively, and $U_{p}=$\ $F_{0}^{2}/(4\omega ^{2})$ is the ponderomotive
energy, which results negligible in the present cases. The width of these
ATI peaks depends on the pulse duration, decreasing as $\tau $ increases
\cite{Faraggi09}, as also observed in PE from atoms \cite{Rodriguez04}.
However, in contrast to the atomic case, the electron emission from metal
surfaces presents a lower limit in the width of the ATI peaks, which is
produced by the energy spread of the metal valence band, characterized by
the Fermi energy. This fact causes the absence of ATI structures in the
multiphotonic spectra for $\omega =$ $0.057$ a.u. (left lower panels of Fig. %
\ref{vsFrequency}) because the energy difference between consecutive ATI
peaks is much lower than the width of each peak.

From Fig. \ref{vsFrequency} it is also observed that the multiphotonic
spectra for $\omega =$ $1.5$ a.u. (right lower panels) show different
structures in the low energy region, depending on the crystallographic
orientation of the surface. The origin of such crystal face effects can be
attributed to two different contributions: the emission from partially
occupied SESs and the induced potential. These contributions will be
separately analyzed in the following subsections.

\subsection{Photoelectron emission from SESs}

\label{ses_section}

\begin{figure}[tbp]
\begin{center}
\includegraphics[width=0.5\textwidth]{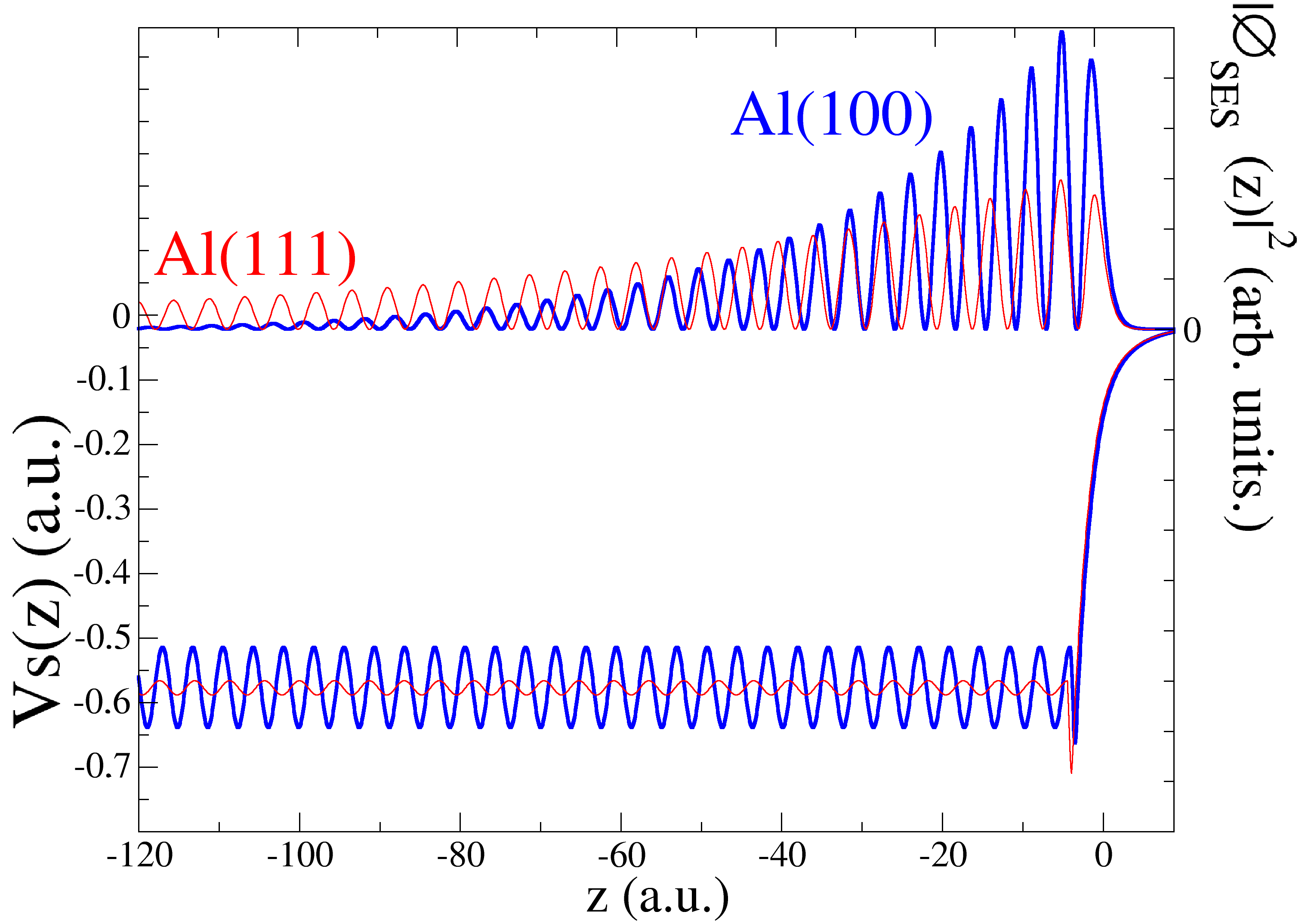}
\end{center}
\caption{(Color online) Potential $V_{S}(z)$ (lower graph), together with
the square modulus of the corresponding SESs, $\left\vert \protect\phi %
_{_{SES}}(z)\right\vert ^{2}$ (upper graph), for Al(100), with blue thick
lines, and for Al(111), with red thin lines. }
\label{potencial-SES}
\end{figure}

One of the most remarkable effects of the crystal band structure is the
presence of partially occupied SESs, which display a highly localized
electron density at the edge of the crystal surface, favoring the release of
electrons from the material. Even though for high carrier frequencies, SESs
were found to play a minor role in PE spectra from Al(111) \cite{BSB-V},\
the relative importance of the SES contribution varies with the
crystallographic orientation and the parameters of the laser pulse. In Fig. %
\ref{potencial-SES}, we plot the surface potential $V_{S}(z)$ for Al(100)
and Al(111), together with the square modulus of the corresponding SESs, $%
\left\vert \phi _{_{SES}}(z)\right\vert ^{2}$, with eigenenergies $%
\varepsilon _{_{SES}}=-0.263$ a.u. and $\varepsilon _{_{SES}}=-0.32$ a.u.,
respectively. The average depth of the potential well, defined as $%
V_{S0}=E_{F}+E_{W}$, is similar for both orientations. However, the
corrugation of $V_{S}(z)$ for the (100) face is almost a factor $6$ larger
than the one corresponding to the (111) orientation, producing a stronger
localization of the SESs of Al(100) close to the surface border, as observed
in Fig. \ref{potencial-SES}.

\begin{figure}[tbp]
{\normalsize \centerline{\includegraphics[width=0.48%
\textwidth]{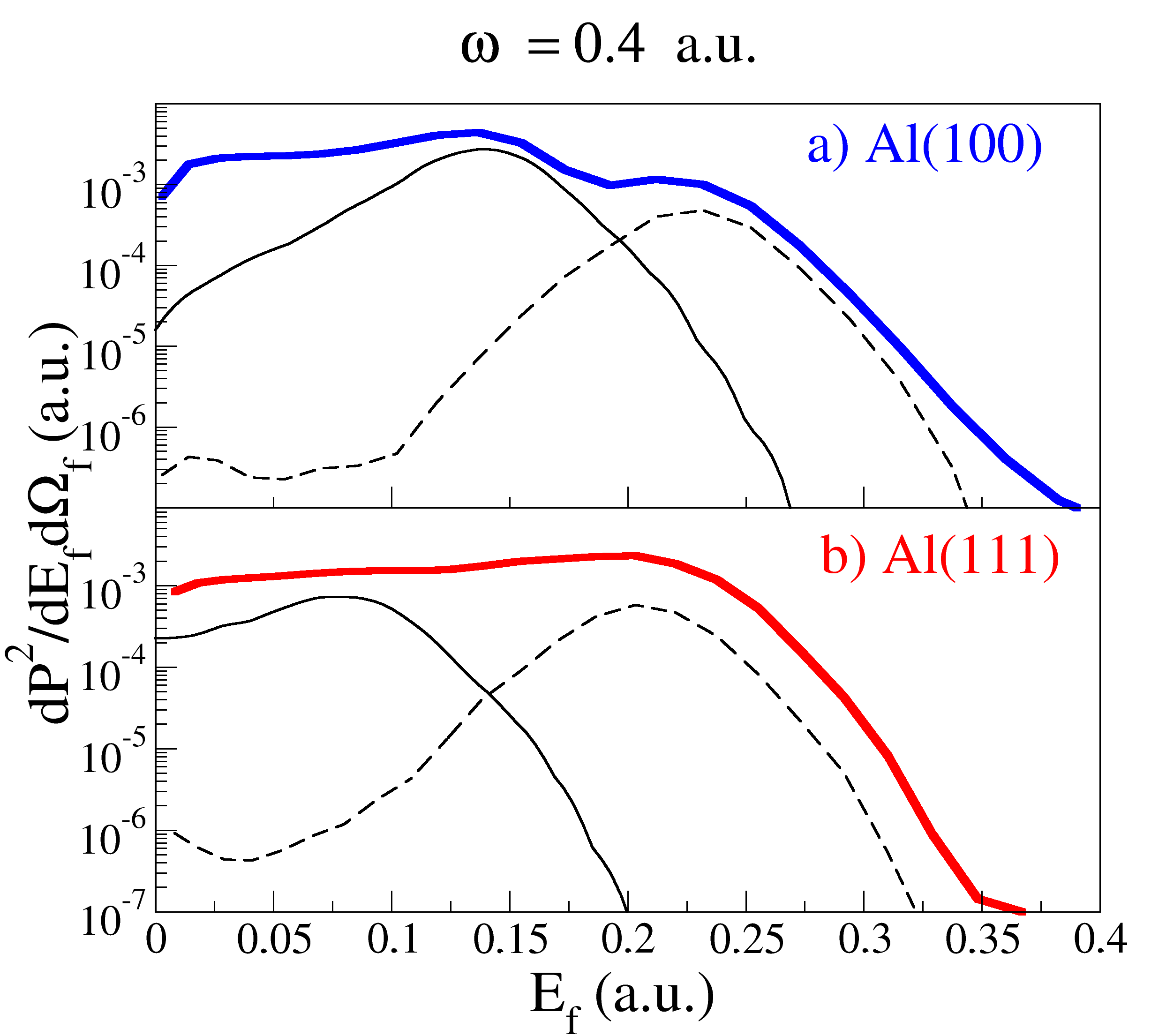}} }
\caption{(Color online) PE distribution in the normal direction, as a
function of the final electron energy, for a 6-cycle laser pulse with a
carrier frequency $\protect\omega =0.4$ a.u. BSB-V results for (a) Al(100)
and (b) Al(111) are displayed in the upper and lower panels, respectively.
In both panels, the thick solid line corresponds to the emission probability
from the valence band, and the black thin dashed and solid lines, to partial
contributions from the top of the valence band and  partially occupied SESs,
respectively. }
\label{vsSES-w04}
\end{figure}

The marked difference between the SES densities of both aluminum faces at
the crystal border can be traced from the superimposed structures of the PE
spectra corresponding to the resonant \ frequency $\omega \simeq \omega
_{s}= $ $0.4$ a.u. For a $6$- cycle laser pulse with $\omega =$ $0.4$ a.u.,
in Fig. \ref{vsSES-w04} we compare differential emission probabilities from
Al(100) and Al(111) with partial values due to emission from SESs as well as
from states at the top of the valence band. 
Despite the general shape of the spectra of Fig. \ref{vsSES-w04} is not
affected by the crystallographic orientation, a closer examination of the
first ATI peak reveals the presence of different superimposed structures for
each face, as also observed in Fig. \ref{vsFrequency}. For Al(100) \ the \
first ATI peak displays a double-bump structure, with two bulges peaking at $%
E_{f}\approx 0.14$ a.u. and $E_{f}\approx 0.24$ a.u. respectively, whereas
for Al(111) only one maximum at $E_{f}\approx 0.24$ a.u. exists. This latter
maximum, present for the two faces, is placed at $E_{f}\simeq $ $%
-E_{W}+\omega $, corresponding to one-photon absorption from initial states
at the top of the valence band, whose partial contribution is also displayed
in the figure. Instead, the peak at $E_{f}\sim 0.14$ a.u., only visible in
the Al(100) spectrum, is produced by the absorption of one-photon from
partially occupied SESs, being placed at $E_{f}\simeq $ $\varepsilon
_{_{SES}}+\omega $. From the comparison of \ Figs. \ref{vsSES-w04} (a) and
(b) we conclude that the presence or absence of SES signatures in the PE
distribution, depending on the crystal face, is associated with the mayor or
minor localization of the electronic density of SESs at the crystal border.
For Al(100) such a localization is more than three times higher than the one
corresponding to Al(111), making its contribution clearly discernible in the
\ PE spectrum. While for Al(111), the \ SES contribution results completed
concealed by emission from other initial states, which washes out SES
footprints from the electron distribution. Therefore, PE spectra under
resonant conditions might offer an attractive window to obtain information
about the surface band structure. Besides, in the resonant case the
contribution of the plasmon decay mechanism should be also included,
producing an additional structure just at the electronic energy $E_{f}\simeq
$ $\omega _{s}$.

\begin{figure}[tbp]
{\normalsize \centerline{\includegraphics[width=0.48%
\textwidth]{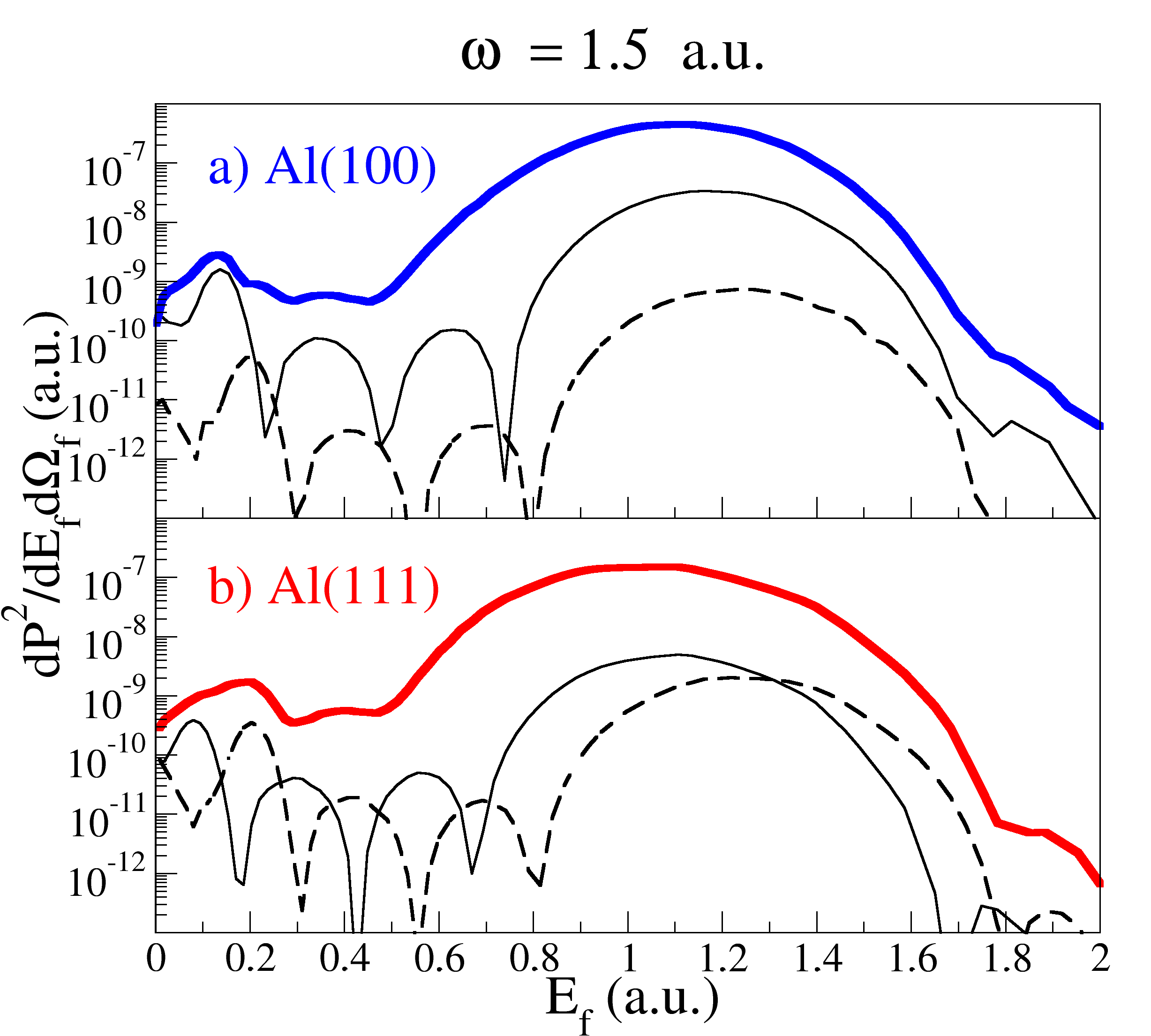}} }
\caption{(Color online) Analogous to Fig. \protect\ref{vsSES-w04} for $%
\protect\omega = 1.5$ a.u. }
\label{vsSES-w15}
\end{figure}

Previous band-structure effects disappear as the carrier frequency departs
from the resonant one. But noteworthy they become again visible for high
frequencies, as shown in Fig. \ref{vsSES-w15}. In this figure we consider
electron emission from Al(100) and Al(111) for a laser pulse with $\omega =$
$1.5$ a.u. and $6$ cycles. In this case, different low-energy structures are
present in the PE spectra for the two faces. The low-energy hump due to
ejection of slow electrons from the top of the valence band is again visible
for both faces, while the one associated with emission from SESs is
appreciable only for the (100) face. As discussed above, this is a
consequence of the different corrugation and SES electronic density of the
two crystallographic orientations. The observation of such effects
significantly depends on the action of the induced potential. Furthermore,
even though for high frequencies these low-energy structures are less
noticeable than in the resonant case, they might still provide information
about the relative importance of SES contributions.

\subsection{Induced potential effects}

\label{induced_section}

\begin{figure}[tbp]
\includegraphics[angle=0, width= 0.49\textwidth]{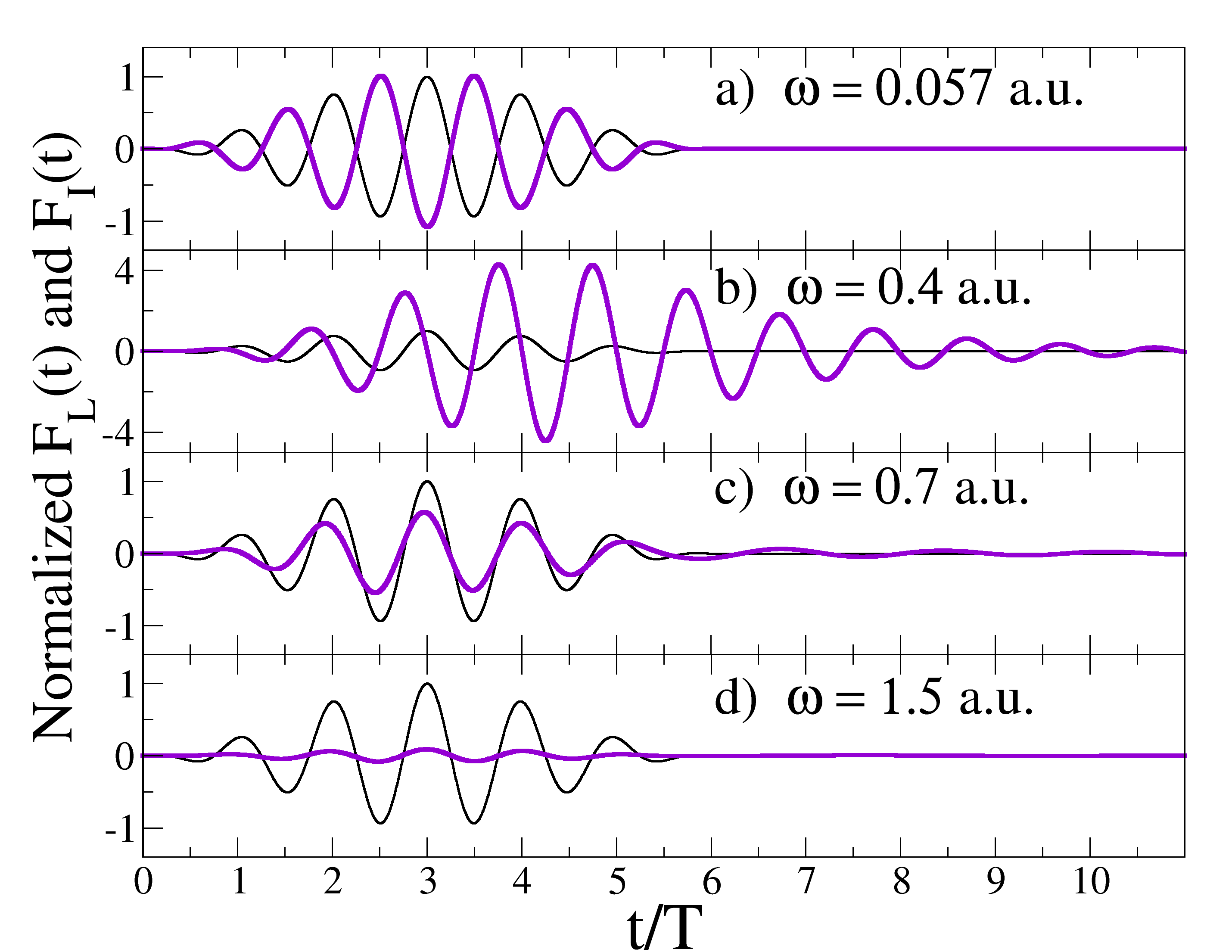}
\caption{(Color online) Normalized laser and induced fields, as a function
of the normalized time $t/T$, for a 6-cycle laser pulse impinging on
Al(111). The carrier frequency of the laser pulse is: a) $\protect\omega %
=0.057$ a.u., b) $\protect\omega =0.4$ a.u. , c) $\protect\omega =0.7$ a.u.,
and d) $\protect\omega =1.5$ a.u. In all panels, black thin line, normalized
laser field $F_{L}(t)/F_{0}$; violet thick line, normalized induced field $%
F_{I}(t)/F_{0}$. }
\label{FI-w}
\end{figure}

\begin{figure*}[tbp]
\includegraphics[width=1.0\textwidth]{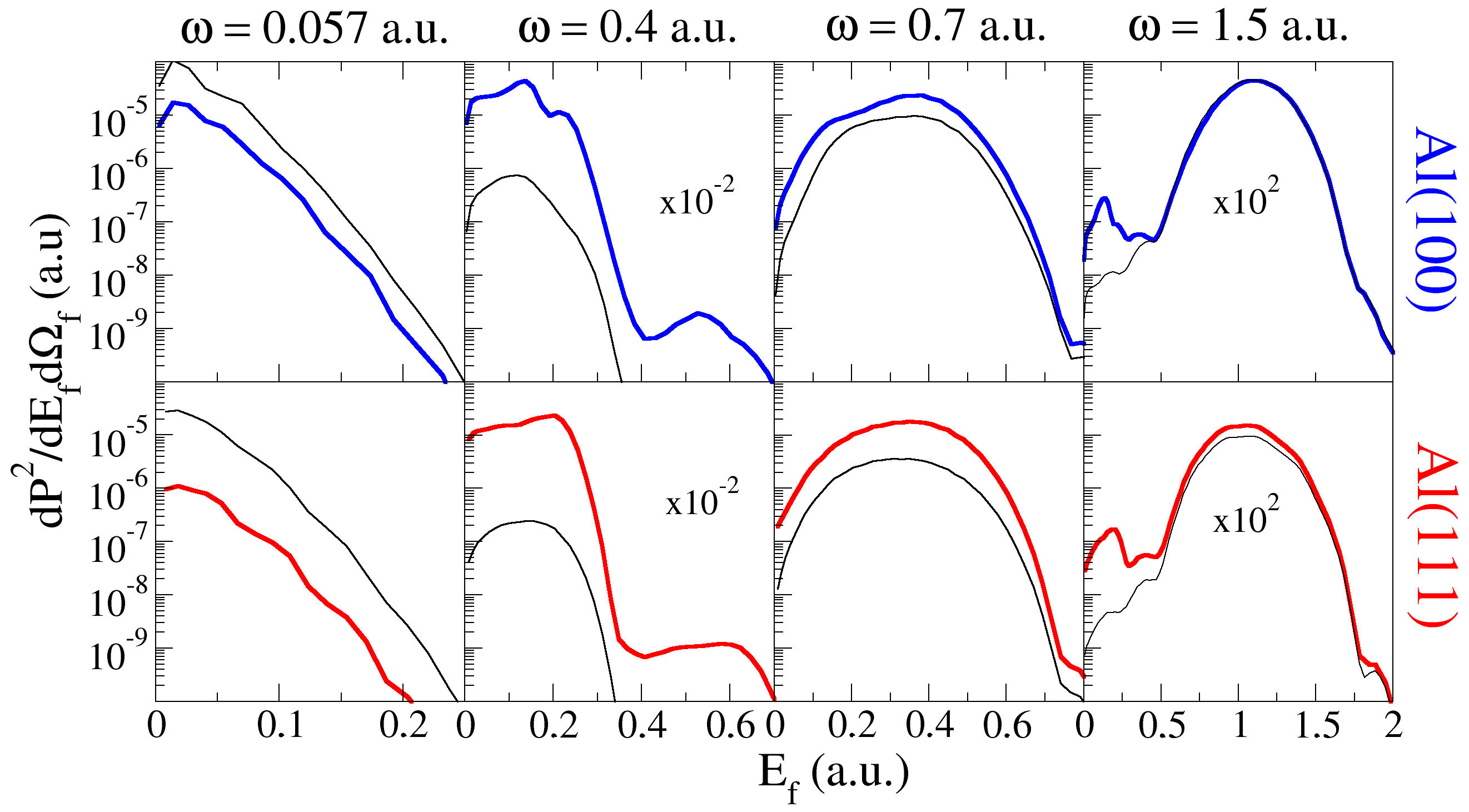}
\caption{(Color online) PE distribution from the valence band of both
aluminum faces - Al(100) (upper row) and Al(111) (lower row) - as a function
of the final electron energy. The impinging 6-cycle laser pulse has a
carrier frequency $\protect\omega =0.057$, $0.4$, $0.7$ and $1.5$ a.u. from
left to right columns, respectively. Thick (thin) solid lines, BSB-V results
obtained including (without including) the induced potential. }
\label{vsInduced}
\end{figure*}

\begin{figure*}[tbp]
\includegraphics[width=1.0\textwidth]{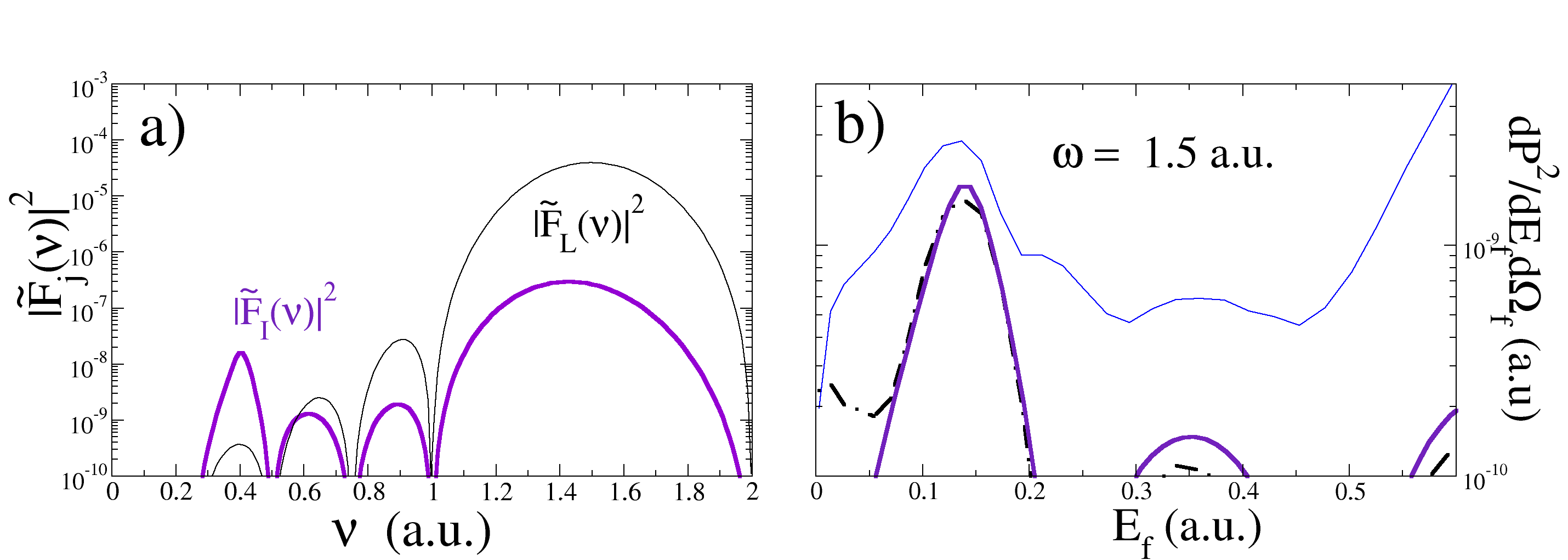} 
\caption{(Color online) Analysis of the frequency domain distribution of
partial contributions to PE from the valence band of Al(100) by a 6-cycle
laser pulse with $\protect\omega = 1.5$ a.u. (a) Frequency profiles of the
laser and induced fields, as a function of the frequency. The Fourier
transform of the laser field $\tilde{F}_{L}$ (induced field $\tilde{F}_{I}$)
is displayed with black thin (violet thick) solid line. (b) The BSB-V PE
distribution, as a function of the electron energy, is shown with blue thin
line, and the SES contribution with black dot-dashed line. The Fourier
transform contribution, $|\tilde{F}_{tot}(E_f+\protect\varepsilon %
_{_{SES}})|^2$ (in arb. units), is plotted with violet solid line. }
\label{TFI-w}
\end{figure*}

With the aim of understanding how the collective response of valence-band
electrons to the laser field affects the distribution of emitted electrons,
in this subsection we examine the contribution of the induced field to PE
spectra.

For laser pulses with several oscillations, the electronic rearrangement
induced in the metal by the external perturbation strongly depends on the
carrier frequency. To illustrate such a variation, in Fig. \ref{FI-w} we
plot the laser and induced fields for $6$-cycle laser pulses, with different
frequencies, impinging on the Al(111) surface. \ For frequencies much lower
than $\omega _{s}$, like $\omega =0.057$ a.u., the induced and external
fields are similar in strength but in counterphase, tending to the static
limit in which the total electric field vanishes inside the metal. But when
the frequency augments to reach the resonant value $\omega \simeq \omega
_{s}=$ $0.4$ a.u., the maximum strength of $F_{I}(t)$ is four times larger
than $F_{0}$; therefore, the total field inside the metal becomes dominated
by the induced response of the metal surface. In turn, for $\omega >\omega
_{s}$ \ surface electrons try to follow the external field, in such a way
that the laser and induced fields oscillate almost with the same phase, but
the intensity of the induced field decreases steeply as the carrier
frequency augments.

The meaningful differences among the induced responses for different
frequencies of the external field, observed in Fig. \ref{FI-w}, are directly
reflected in their contributions to the electron emission process. In Fig. %
\ref{vsInduced} we compare BSB-V differential emission probabilities,
derived from Eq. (\ref{dPdEdW}), with BSB-V values obtained without
including the induced surface interaction, that is, by fixing $%
F_{I}(t)=A_{I}(t)=0$, considering the same laser parameters as in Fig. \ref%
{FI-w}. Results for both crystallographic orientations -Al(100) (upper
panels) and Al(111) (lower panels)- are displayed in the figure. From Fig. %
\ref{FI-w} (a), for $\omega =0.057$ a.u. 
the induced response 
screens the laser field inside the metal almost completely.
Then, in this case the
inclusion of the induced potential causes a marked reduction of the PE
yield, this effect being stronger for Al(111) than for Al(100). Instead, for
$\omega =0.4$ a.u. the laser pulse becomes resonant with $\omega _{s}$,
originating a large increase of the emission probability, greater than two
order of magnitude. Notice that for this resonant frequency, the induced
field is not only higher than the external field but also it persists twice
the original pulse duration (see Fig. \ref{FI-w} (b)), contributing to
augment the emission probability after the external field turned off. For a
carrier frequency slightly higher than the surface plasmon frequency, like $%
\omega =0.7$ a.u., the induced field goes on reinforcing the action of the
laser pulse, although its intensity is four times lower than that
corresponding to the resonant case, as shown in Fig. \ref{FI-w} (c). Hence,
the inclusion of induced potential within the BSB-V approach produces a
moderate increase of the PE probability, being again higher for the (111)
face.

The case of $\omega =1.5$ a.u., shown in the right column of Fig. \ref%
{vsInduced}, deserves further discussion. For such a high frequency,
electrons are not able to follow the quick variation of the external
perturbation and consequently, the maximum strength of $F_{I}(t)$ is more
than one order of magnitude lower than the one of the laser field, as
displayed in Fig. \ref{FI-w} (d). As it was expected, this small induced
response does not affect appreciably the main electron emission, which
occurs around the first ATI peak. However, we remarkably found that the
induced potential introduces a pronounced growth of the probability at low
electron energies, just in the region where the double-hump low-energy
structure associated with SES emission appears in the Al(100) case. To
investigate in detail this unforeseen contribution, in Fig. \ref{TFI-w} (a)
we plot the decomposition in frequencies of both the laser ($\tilde{F_{L}}%
(\nu )$) and the induced field ($\tilde{F_{I}}(\nu )$) for this case. The
utility of analyzing the frequency domain of the fields lies on the fact
that the PE spectrum can be roughly estimated as proportional to the square
modulus of the Fourier transform of the total electric field, that is, $|%
\tilde{F}_{tot}(\nu )|^{2}$, evaluated at $\nu =E_{f}+E_{i}$, with $\tilde{F}%
_{tot}(\nu )=\tilde{F}_{L}(\nu )+\tilde{F}_{I}(\nu )$ and $E_{i}$ covering
the energy range of all initially occupied states. In Fig. \ref{TFI-w} (a),
although the Fourier transform of the laser field is several orders of
magnitude higher than the one of the induced field around the carrier
frequency, \ $\tilde{F_{I}}(\nu )$ retains a peak associated with the
resonance $\nu \cong \omega _{s}$, which largely overpasses the value of $%
\tilde{F}_{L}(\omega _{s})$. This resonant peak is found to be the origin of
the low-energy SES contribution of the spectrum for Al(100), as shown in
Fig. \ref{TFI-w} (b), where we observe that the curve corresponding to $|%
\tilde{F}_{tot}(E_{f}+\varepsilon _{_{SES}})|^{2}$ (multiplied by an
arbitrary factor) almost coincides with the SES contribution.

\section{Conclusions}

We have studied PE spectra produced by the interaction of ultrashort laser
pulses with the valence band of \ two different faces of aluminium - Al(100)
- Al(111) - by using of the BSB-V approximation. In the present version of
the BSB-V approach we have incorporated the contribution of the induced
field, originated from the collective response of surface electrons to the
external perturbation. We found that the induced response of the metal
surface strongly affects electron emission distributions for a wide range of
laser frequencies, including high $\omega $- values. In the resonant case
with the surface plasmon frequency,  the effect of the induced field
contributes to make visible signatures coming from partially occupied SESs
of the Al(100) surface, while for Al(111) these SES structures are
completely washed out by emission from other initially occupied states.
Similar features are also observed in the low-energy region of the spectra
for high carrier frequencies of the laser pulse, for which the influence of
the induced potential was expected negligible. These findings open the way
to investigate band structure effects by varying the parameters of the laser
pulse, the crystal orientation, and the observation region.



\begin{acknowledgments}
Financial support from CONICET, UBA, and ANPCyT of Argentina is acknowledged.
\end{acknowledgments}

\bigskip \bigskip

\bibliographystyle{unsrt}
\bibliography{ref}

\end{document}